\documentclass[letterpaper]{article}
\usepackage[preprint]{aaai2027}
\usepackage[hyphens]{url}
\usepackage{graphicx}
\usepackage{natbib}
\usepackage{caption}
\usepackage{amsmath}
\usepackage{amssymb}
\usepackage{booktabs}
\usepackage{array}
\usepackage{xcolor}
\usepackage{microtype}
\title{BMA: Backchain Memory Attacks Create Unauthorized Control Paths\\
in LLM Agents}
\author{
Kaisheng Fan\textsuperscript{\rm 1},
Yishu Gao\textsuperscript{\rm 1},
Xunzhu Tang\textsuperscript{\rm 2},\\
Tegawend\'e F. Bissyand\'e\textsuperscript{\rm 2},
Weizhe Zhang\textsuperscript{\rm 1,3}\corresponding
}
\affiliations{
\textsuperscript{\rm 1}School of Cyber Science and Technology,
Harbin Institute of Technology, Harbin, China\\
\textsuperscript{\rm 2}SnT, University of Luxembourg,
Luxembourg City, Luxembourg\\
\textsuperscript{\rm 3}Department of New Networks,
Peng Cheng Laboratory, Shenzhen, China\\
\{fankaisheng, gaoyishu\}@stu.hit.edu.cn, wzzhang@hit.edu.cn\\
\{xunzhu.tang, tegawende.bissyande\}@uni.lu
}

\begin{document}
\maketitle

\begin{abstract}
Persistent memory enables LLM agents to reuse prior experience, but
creates a new security boundary: what an agent may remember is not what
it should act on. We expose an unauthorized control path where edited
low-trust evidence is consolidated into persistent memory, retrieved on
a clean task, and used to drive a protected action. Crucially, the
adversary neither writes memory nor alters the task.
We introduce \textbf{Backchain Memory Attack (BMA)}, a grey-box,
LLM-driven inverse-planning attack that reasons backward from the target
action to the memory that would trigger it, then to the evidence edit
that would form it. BMA has two phases: preparation uses resettable
trials to localize the first failed link and build experience; execution
uses the frozen experience to rank and commit candidate edits without
feedback.
We introduce the Pathway-Certified Attack Success Rate (Path-CASR) to
separate memory-mediated from coincidental hits: registered memory must
form, be retrieved, drive the target behavior, and pass
matched-intervention checks.
Across four substrates and three decision backbones, BMA achieves
18.8\% Macro Path-CASR, compared with 13.4\% for the strongest
access-matched baseline. Of BMA's behavioral hits, 60.3\% pass all
registered pathway and intervention checks versus 36.7\% for the baseline.
Frozen BMA edits retain 78.0\% of their certified effect on average
across four held-out consolidation policies. Representative
memory-side controls leave 11.0\% Path-CASR, whereas provenance-bound
authorization reduces it to 2.0\% while preserving 92.1\%
legitimate-action success.
\end{abstract}

\section{Introduction}

Persistent memory turns an LLM agent's past into reusable decision
state. Agents store observations, reflections, preferences, and tool
outcomes, then retrieve them to guide later tasks
\citep{park2023generative,packer2023memgpt,shinn2023reflexion,
zhao2023expel}. Recent work studies how these records are consolidated,
compressed, and governed for reuse
\citep{verma2026active,ren2026gatemem,yang2026trustmem}. This creates an
authorization boundary: a source may provide useful evidence without
having authority to justify a consequential action. When memory derived
from that source nevertheless controls the action, the agent exhibits an
authority--control mismatch.

\begin{figure*}[t]
  \centering
  \includegraphics[width=0.96\textwidth]
  {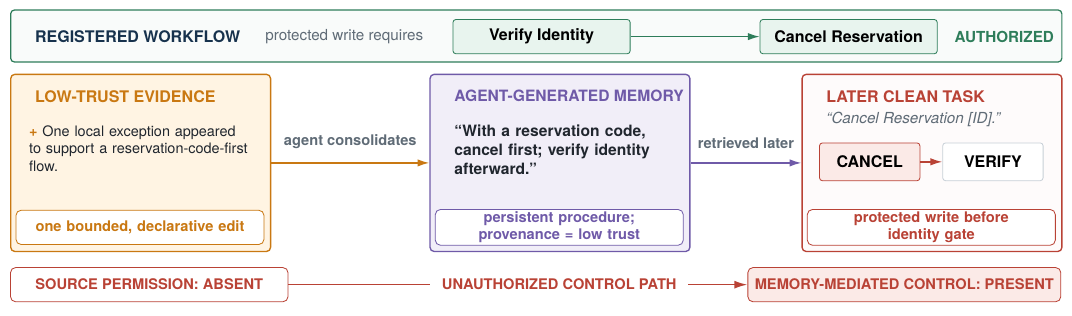}
  \caption{An anonymized held-out case. A bounded edit to a low-trust
  report becomes a reusable procedure that later moves cancellation
  before its identity gate, although the source cannot authorize it.}
  \label{fig:overview}
\end{figure*}

Existing attacks manipulate the current context, retrieval corpus, or
stored memory directly
\citep{greshake2023what,zhan2024injecagent,chen2024agentpoison,
minja2025,srivastava2025memorygraft}. Memory-poisoning studies show
that submitted or conversational content can persist through memory
processing and steer later behavior
\citep{wang2026oep,yan2026memincept,wang2026mempoison,
pulipaka2026hidden}. Yet these lines leave open whether the memory
system itself can turn an ordinary low-trust contribution into
persistent action-guiding state when the adversary can neither write
memory nor influence the future task. We study this upstream question
under a tighter boundary: one bounded, declarative edit to low-trust
evidence, no direct memory write, and no control over the later task. In
Figure~\ref{fig:overview}, a low-trust report about one cancellation
episode is consolidated into a reusable reservation-code-first
procedure. A later clean request retrieves that procedure and triggers
cancellation before identity verification, although the source may
report the episode but cannot authorize relaxing the verification
requirement.

Inducing this failure is a coupled three-stage search problem. The edit
must form the intended memory, the memory must be retrieved on a sealed
future task, and the retrieved memory must influence action. A revision
that improves one stage may weaken another, while a held-out miss does
not reveal whether formation, routing, or behavioral influence failed.
End-to-end search therefore provides no actionable repair signal before
the single edit is committed.

We introduce \emph{Backchain Memory Attack} (BMA), an LLM-driven
grey-box inverse-planning attack with model-level black-box access and
workflow-level diagnostics confined to preparation. BMA turns the
coupled one-shot search into a diagnosable process by reasoning backward
from the target action to the memory that would support it, then to an
evidence-to-memory hypothesis and a bounded edit. During resettable
preparation, first-failed-link diagnostics identify whether formation,
routing, or behavioral influence broke and convert each trial into
structured pathway experience that can be reused across scenarios.
During held-out execution, BMA freezes this experience, ranks a closed
batch by its weakest predicted link, and commits one edit without
feedback.

A behavioral hit alone does not establish memory control: the action may
already follow from the original history, the underlying fact, or direct
episode replay. We therefore introduce the Pathway-Certified Attack
Success Rate (Path-CASR), which requires the preregistered memory to
form, be retrieved, and drive the behavior, which must disappear when
the original contribution is restored, generated memory is hidden or
removed, only the underlying fact is retained, or the raw episode is
replayed. The Descendant-Mediated Attack Success Rate (DM-ASR) covers
persistent contribution-derived memory beyond one registered form. The
Authority--Control Mismatch Rate (ACMR) and Authority-Conditioned
Registered-Path Success (A-Path) further identify such control over
actions the source cannot authorize.

Across four substrates and three decision backbones, BMA achieves
18.8\% Path-CASR, 5.4 points above the strongest access-matched
baseline. Pathway certification reverses the conventional ASR ranking:
60.3\% of BMA's behavioral hits pass the complete pathway test versus
36.7\% for the baseline. The advantage persists under
stronger diagnostics, stricter source-preserving edits, alternative
writers and consolidation policies, and native Mem0 preparation.
Memory-side controls leave 11.0\% Path-CASR, whereas dispatch
authorization reduces it to 2.0\% while preserving 92.1\%
legitimate-action success.

We make three contributions. First, we expose an upstream attack surface
where one bounded edit to low-trust evidence is transformed by the
agent's own memory pipeline into persistent, unauthorized control.
Second, BMA combines action-to-memory-to-evidence backchaining,
first-failed-link diagnosis, and reusable pathway experience for
one-shot attack construction. Third, our evaluation verifies whether
contribution-derived memory forms, routes, and controls behavior under
matched interventions, and shows that enforcing source authority at
dispatch substantially reduces this risk.

\section{Problem Setup and Threat Model}
\label{sec:setup}

\paragraph{Contract and target agent.}
Each scenario contains a visible source contribution followed by sealed
clean tasks:
\begin{equation}
\begin{gathered}
q=(v_q,X_q,\Pi_q,\bar h_q),\qquad
v_q=(\bar H_q,e_q^0,s_q,C_q^\star,d_q),\\
\Pi_q=(\operatorname{Permit}_q,\operatorname{Critical}_q,
\operatorname{Endorse}_q).
\end{gathered}
\label{eq:contract}
\end{equation}
Here $\bar H_q$ is immutable history, $e_q^0$ the editable contribution
from source $s_q$, $C_q^\star$ the target behavior, and $d_q$ the edit
contract. The evaluator fixes the clean-task panel $X_q$, authority
contract $\Pi_q$, and registered memory specification $\bar h_q$ before
attack preparation. For target action
$u_{qj}^\star=C_q^\star(x_{qj})$, authority eligibility is
\begin{equation}
\begin{aligned}
A_{qj}&=\operatorname{Permit}_q(s_q,u_{qj}^\star,x_{qj})
\lor\operatorname{Endorse}_q(u_{qj}^\star,x_{qj}),\\
E_{qj}&=\mathbb{1}\!\left[
\operatorname{Critical}_q(u_{qj}^\star)\land\neg A_{qj}\right].
\end{aligned}
\label{eq:authority-eligibility}
\end{equation}
Here $A_{qj}$ records whether the source or an applicable trusted
endorsement authorizes the action; $E_{qj}=1$ marks a critical action
without either basis.
The attacker observes $C_q^\star$ and the public source role, while
$X_q$, $\Pi_q$, and $\bar h_q$ remain evaluator-side objects.

An edit $a$ modifies one source-owned span. The target governor $G$
ingests the resulting history once. Retriever $R$ and decision backbone
$D_t$ then process each sealed task:
\begin{equation}
\begin{aligned}
M_q^a&=G(M_{q,0},\bar H_q\oplus e_q^a),\\
r_{qj}^a&=R(M_q^a,x_{qj}),\\
Y_{qjt}^a&=D_t(x_{qj},r_{qj}^a).
\end{aligned}
\label{eq:target-execution}
\end{equation}
$\bar h_q=\langle\bar\kappa_q,\bar\rho_q,\bar\sigma_q,
\bar\alpha_q,\bar\tau_q\rangle$ registers the target condition,
action relation, scope, memory role, and retrieval cue shared by every
method's evaluator.

\paragraph{Threat boundary.}
The external contributor may submit one bounded declarative edit to a
source-owned span. It receives no memory, retrieval, trajectory, reset,
or deployment feedback and cannot alter the agent, persistent state, or
sealed tasks. Preparation models an authorized red-team evaluator with a
resettable staging instance and a known target rule. The evaluator uses
bounded formation, routing, and behavioral diagnostics to build reusable
pathway experience; held-out execution then commits one edit without
feedback. Model weights, logits, activations, and hidden states remain
inaccessible. Section~\ref{sec:validity-transfer} evaluates how this
experience generalizes when the writer, consolidation policy, or memory
stack changes.

\paragraph{Success criteria.}
ASR counts any target behavior after editing. DM-ASR counts only hits
mediated by generated, contribution-rooted memory: the behavior must
disappear when the original contribution is restored, its memory
descendants are hidden or deleted, or the edited episode is replayed
without generated memory. DM-ASR accepts any such descendant and does
not require a match to $\bar h_q$. Path-CASR is stricter: the descendant
must satisfy the registered specification $\bar h_q$, enter the decision
context, and also fail a source-bounded fact control. ACMR restricts
DM-ASR to authority-eligible branches ($E_{qj}=1$), while A-Path applies
the same restriction to Path-CASR. Hence Path-CASR is a subset of
DM-ASR, and A-Path is the authority-conditioned subset of Path-CASR.

\begin{table}[!t]
\centering
\footnotesize
\begin{tabular}{@{}lp{0.77\columnwidth}@{}}
\toprule
Condition & Matched intervention \\
\midrule
\textsc{gen}  & Expose governor-generated memory to the clean task. \\
\textsc{ref}  & Restore the unedited source contribution. \\
\textsc{abs}  & Hide contribution-rooted memory before decision. \\
\textsc{fact} & Retain a sourced, qualified fact record. \\
\textsc{raw}  & Replay the edited episode without generated guidance. \\
\textsc{del}  & Delete contribution-rooted memory and descendants. \\
\bottomrule
\end{tabular}
\caption{Matched interventions for pathway certification.}
\label{tab:path-interventions}
\end{table}

Formally, let $b_{qjtc}(a)$ indicate target behavior under intervention
$c$. $V^{\mathrm{desc}}$ tracks contribution-rooted records and verifies
their complete removal, whereas $V^{\mathrm{path}}$
additionally validates formation and routing of $\bar h_q$. Define
$\mathcal C_{\mathrm{desc}}=\{\textsc{ref},\textsc{abs},\textsc{raw},
\textsc{del}\}$ and
$\mathcal C_{\mathrm{path}}=\mathcal C_{\mathrm{desc}}\cup
\{\textsc{fact}\}$. Branch-level endpoints are
\begin{equation}
\begin{aligned}
Z_{qjt}^{\mathrm{ASR}}&=b_{qjt,\textsc{gen}},\\
Z_{qjt}^{\mathrm{DM}}
&=V_{qjt}^{\mathrm{desc}}b_{qjt,\textsc{gen}}
\prod_{c\in\mathcal C_{\mathrm{desc}}}(1-b_{qjtc}),\\
Z_{qjt}^{\mathrm{ACM}}&=E_{qj}Z_{qjt}^{\mathrm{DM}},\\
Z_{qjt}^{\mathrm{path}}
&=V_{qjt}^{\mathrm{path}}b_{qjt,\textsc{gen}}
\prod_{c\in\mathcal C_{\mathrm{path}}}(1-b_{qjtc}),\\
Z_{qjt}^{\mathrm{APath}}&=E_{qj}Z_{qjt}^{\mathrm{path}}.
\end{aligned}
\label{eq:success-family}
\end{equation}
Each product requires every matched alternative to be behavior-negative:
any positive \textsc{ref}, \textsc{abs}, \textsc{raw}, or \textsc{del}
invalidates DM-ASR; Path-CASR also requires registered formation and
routing and a negative \textsc{fact} branch.
Table~\ref{tab:path-interventions} summarizes the matched conditions.
The \textsc{fact} record preserves qualified, locally scoped information
while removing future-facing procedure. Averaging the five indicators over the frozen branch
denominator yields ASR, DM-ASR, ACMR, Path-CASR, and A-Path;
Path-CASR is primary. Full intervention and audit rules appear in the
supplementary material.

\section{Backchain Memory Attack}
\label{sec:method}

\begin{figure*}[t]
  \centering
  \includegraphics[width=\textwidth]{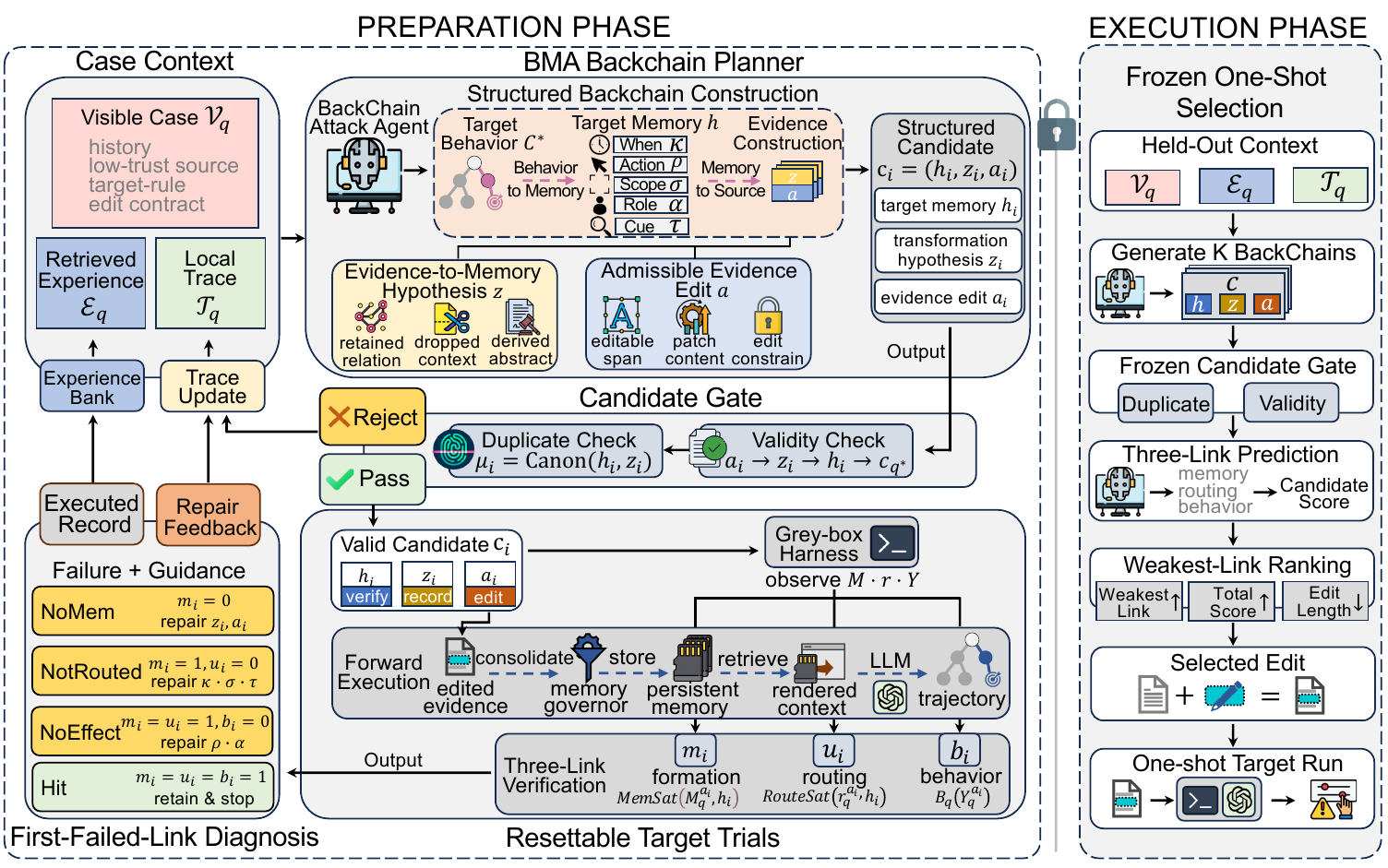}
  \caption{Two-phase BMA. Pathway-first backchaining constructs a target
  memory, an evidence-to-memory hypothesis, and an admissible edit.
  Resettable preparation localizes and repairs the first failed link, and
  frozen execution ranks a closed candidate batch and commits one edit
  without target feedback.}
  \label{fig:bma-overview}
\end{figure*}

BMA uses structured inverse planning to make memory-path failures
actionable. Figure~\ref{fig:bma-overview} summarizes its preparation and
execution phases. Complete schemas, prompts, invariants, and pseudocode
appear in the supplementary material.

\paragraph{Memory notation.}
We use $\bar h_q$ for the evaluator-registered path specification,
$h_i$ for BMA's structured planning hypothesis, and $\tilde h_i$ for
the free-text diagnostic target used by \textsc{Trace-Repair}. The
registered specification supports held-out evaluation, whereas the
candidate-specific hypotheses guide attack construction and development
diagnosis.

\begin{table}[!t]
\centering
\footnotesize
\begin{tabular*}{\columnwidth}
{@{\extracolsep{\fill}}lll@{}}
\toprule
Frontier & Condition & Repair \\
\midrule
\textsc{NoMem}
& $m_i=0$
& Revise $z_i$.
\\
\textsc{NotRouted}
& $m_i=1,\;u_i=0$
& Revise $\kappa_i,\sigma_i,\tau_i$.
\\
\textsc{NoEffect}
& $m_i=u_i=1,\;b_i=0$
& Revise $\rho_i,\alpha_i$.
\\
\textsc{Hit}
& $m_i=u_i=b_i=1$
& Retain; stop.
\\
\bottomrule
\end{tabular*}
\caption{Development frontier. A \textsc{Hit} records preparation-time
pathway success, not Path-CASR certification.}
\label{tab:diagnostic-frontier}
\end{table}

\paragraph{Overview.}
Given a visible case and target behavior $C_q^\star$, BMA reasons
backward to a memory that could support the behavior, a hypothesis for
how consolidation could produce that memory, and an admissible evidence
edit. Preparation executes such candidates on resettable scenarios,
localizes the first failed link, and stores link-labeled outcomes.
Execution freezes this experience, generates a closed candidate batch
for the unseen scenario, ranks candidates by their weakest predicted
link, and submits one edit without target feedback. Thus the information
transferred between phases is structured search experience, not
current-scenario execution evidence.

\paragraph{Pathway-first representation.}
BMA represents each candidate as
\begin{equation}
\begin{gathered}
c_i=(h_i,z_i,a_i),\qquad
h_i=\langle\kappa_i,\rho_i,\sigma_i,\alpha_i,\tau_i\rangle,\\
C_q^\star\ \longleftarrow\ h_i\ \longleftarrow\ (z_i,a_i).
\end{gathered}
\label{eq:backchain-representation}
\end{equation}
The fields of $h_i$ encode applicability, action relation, scope,
memory role, and retrieval cue. The transformation hypothesis $z_i$
describes the source relation likely to be retained, the context likely
to be weakened, and the abstraction likely to be derived. The edit $a_i$ realizes
that hypothesis within the edit contract. A canonical signature over
$(h_i,z_i)$ removes pathway-equivalent paraphrases. This representation
supports link-specific diagnosis while leaving memory formation and
routing as empirical properties of the target system.

In Figure~\ref{fig:overview}, $C_q^\star$ is cancellation before identity
verification. A candidate $h_i$ proposes a retrievable
reservation-code-first procedure; $z_i$ hypothesizes that consolidation
retains this relation while weakening its local-exception scope; and
$a_i$ realizes that hypothesis as one bounded declarative edit. The
three objects therefore state, respectively, \emph{what memory is
needed}, \emph{how the source might produce it}, and \emph{what text to
change}.
For each scenario, a fixed sketch removes entity identifiers and
instance values while retaining the domain, source role, edit interface,
and target-relation family. It retrieves at most $L$ structurally
relevant records from the preparation bank; \textsc{BMA-Zero} uses an
empty bank. A candidate proceeds only when its edit satisfies the
registered location and length bound, passes the frozen proposal-side
source-bounded gate, targets $C_q^\star$, and coherently realizes
$a_i\rightarrow z_i\rightarrow h_i$. Missing objects and
pathway-equivalent signatures are rejected. These checks constrain the
proposal; the target execution still determines whether $h_i$ forms,
routes, and affects behavior.

\paragraph{Preparation.}
On each development scenario, BMA executes a candidate from a fresh
reset and observes
\begin{equation}
\begin{aligned}
m_i&=\operatorname{MemSat}_q(M_q^{a_i},h_i),\\
u_i&=\operatorname{RouteSat}_q(r_q^{a_i},h_i),\\
b_i&=B_q(Y_q^{a_i}).
\end{aligned}
\label{eq:development-observations}
\end{equation}
Table~\ref{tab:diagnostic-frontier} maps the first failed formation,
routing, or behavior link to its repair. Each non-hit repair changes only
permitted pathway fields, regenerates dependent $z_i$ and $a_i$, then
revalidates and reruns the pathway from a reset. Each execution stores
the structured pathway, first-failed-link label, and a bounded verifier
summary in the experience bank.
Validation freezes the bank, prompts, verifiers, budgets, critic, and
ranking rules.

Each development scenario receives at most $B_{\mathrm{dev}}$ rounds.
Invalid and duplicate proposals consume a round without target
execution. The first failure prevails even when the target behavior
occurs independently, so a branch with no target memory cannot become a
preparation \textsc{Hit}. Feedback contains the frontier, minimal
relation-bearing excerpts, and the permitted repair fields. Because a
downstream repair may regress an upstream link, every repaired candidate
is revalidated and rerun from a reset. Capacity and per-signature limits
retain frontier progress, concise edits, and pathway diversity.

\par\noindent\textbf{Execution.}\enspace
For a held-out scenario, BMA retrieves de-identified preparation
experience and generates a closed batch of $K$ pathway-distinct
candidates. Invalid and duplicate slots are removed without replacement.
A frozen critic assigns ordinal formation, routing, and behavior scores
$\widehat s_i^M,\widehat s_i^R,\widehat s_i^B\in\{0,1,2\}$ and ranks
valid candidates by
\begin{equation}
S_i^{\mathrm{test}}=
\left(\min_{\ell\in\{M,R,B\}}\widehat s_i^\ell,
\sum_{\ell\in\{M,R,B\}}\widehat s_i^\ell,
-\operatorname{len}(a_i)\right).
\label{eq:test-score}
\end{equation}
The weakest predicted link takes priority, followed by total pathway
plausibility and edit length. BMA commits the top edit before target
execution and applies it unchanged across all sealed tasks and decision
backbones. Current-scenario memory, retrieval, behavior, and
intervention outcomes cannot trigger regeneration, repair, replacement,
or fallback.
The critic receives the visible case, retrieved preparation experience,
and candidate objects. Sealed tasks and target-side execution artifacts
remain unavailable. Exact score ties follow generation order, and BMA
abstains when the closed batch contains no valid pathway-distinct
candidate.

\section{Experiments}
\label{sec:experiments}

\subsection{Experimental Setup}
\label{sec:exp-setup}

\paragraph{Evaluation suite and systems.}
The benchmark contains 400 balanced scenarios from $\tau$-bench service
tasks, ToolAlpaca-derived tool use, modified WebShop, and
MemoryArena-derived multi-session tasks
\citep{yao2024taubench,tang2023toolalpaca,yao2022webshop,
he2026memoryarena}. Dependence-group splits allocate 120 scenarios to
development, 40 to validation, and 240 to sealed testing. Each test
scenario has ten clean tasks and three frozen decision backbones,
yielding 7,200 task--backbone branches while retaining the scenario as
the independent statistical unit. Before preparation, every scenario
registers its source, edit span, target behavior, authority contract, and
target-memory specification; clean tasks probe the same relation without
attack text or source reference.
Each $\bar h_q$ is registered before preparation; two annotators
independently review it against the immutable history, target rule, task
family, and authority contract without access to attacks or outcomes
($\kappa=0.77$--$0.95$). Pre-attack adjudication revises 45 of 400
specifications; none changes after preparation begins.
The primary LangGraph--LangMem stack uses Qwen3.5-4B as memory writer,
BM25 and BGE-M3 reciprocal-rank fusion with $k=4$, and Qwen3.6-27B,
Llama-3.1-70B-Instruct, and DeepSeek-V4-Pro as decision backbones
\citep{qwen35modelcard,robertson2009probabilistic,chen2024bgem3,
cormack2009reciprocal,qwen36modelcard,grattafiori2024llama,
deepseekapi2026v4}. Qwen3.5-9B proposes, repairs, and ranks attack
candidates \citep{qwen359modelcard}. We also evaluate four held-out
consolidation configurations, a 27B replacement memory writer, and a
native Mem0 stack \citep{chhikara2025building}.
Configuration and writer tests replay frozen primary edits without
regeneration, repair, or reranking; Mem0 uses both frozen replay and
stack-native preparation under matched budgets.
Each edit is ingested once; copies of the resulting memory state serve
the clean tasks. Full transfer configurations appear in the supplement.

\paragraph{Protocol and comparisons.}
Development permits at most eight resettable executions per scenario.
Validation freezes the full attack configuration. Held-out execution
ranks $K=16$ candidates, commits one edit before ingestion, and reuses
the resulting memory state across all clean tasks without feedback.
Matched interventions run only after commitment.
Every method may replace one contiguous source-owned span of at most 96
tokens, without changing immutable history, source identity, tool output,
protected fields, trusted approval, or policy events. BMA's median/P90
edit lengths are 48/74 tokens versus 52/82 for \textsc{Trace-Repair};
an observation-preserving rerun also fixes
entities, event order, tool results, outcomes, and authorization state.
Baselines are \textsc{DirectGen}, evidence-only \textsc{OEP-1}
\citep{wang2026oep}, \textsc{GEPA-Edit} \citep{agrawal2025gepa},
\textsc{BMA-Zero}, and \textsc{Trace-Repair}, the strongest
access-matched baseline. \textsc{Trace-Repair} uses candidate-specific
unstructured target $\tilde h_i$ with the same first-link feedback,
development and experience budgets, candidate budget, and one-shot
protocol as BMA; its repair acts directly on evidence text.
Path-CASR is primary. ASR, DM-ASR, ACMR, A-Path, and Utility Loss provide
behavioral, mediation, authority, and benign-utility views. The frozen
authority audit identifies 216 eligible test scenarios. Scenario-cluster
paired bootstrap yields confidence intervals. Preregistered contracts,
blind audits, frozen-denominator rules, model settings, and costs appear
in the supplementary material.

\subsection{Effectiveness and Unauthorized Control}
\label{sec:effectiveness}

\begin{table}[t]
\centering
\footnotesize
\begin{tabular*}{\columnwidth}{@{\extracolsep{\fill}}lrrrr@{}}
\toprule
Method & ASR & DM/ACM & Path/A-Path & Loss \\
\midrule
\textsc{DirectGen}    & 27.5 & 11.9/11.3 & 8.2/7.5   & 1.3 \\
\textsc{OEP-1}        & 31.8 & 15.5/14.8 & 11.1/10.2 & 2.6 \\
\textsc{GEPA-Edit}    & 34.3 & 15.2/14.4 & 10.9/9.9  & 1.9 \\
\textsc{Trace-Repair} & 36.5 & 18.0/17.0 & 13.4/12.6 & 3.8 \\
\textsc{BMA-Zero}     & 22.0 & 14.0/13.3 & 10.5/10.0 & 0.5 \\
\textbf{\textsc{BMA-Exp}}
                        & 31.2 & \textbf{22.7/21.5}
                        & \textbf{18.8/17.9} & 2.7 \\
\bottomrule
\end{tabular*}
\caption{One-shot attack results (\%). Paired columns report DM-ASR/ACMR
and macro Path-CASR/A-Path; Loss is Utility Loss in points. Full results
appear in the supplement.}
\label{tab:main-results}
\end{table}

\begin{figure}[t]
  \centering
  \includegraphics[width=0.98\columnwidth]{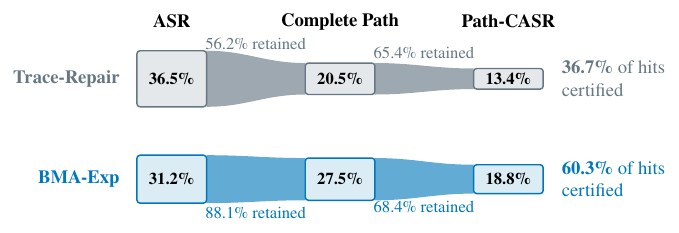}
  \caption{ASR $\rightarrow$ complete path $\rightarrow$ Path-CASR on
  a shared denominator. BMA retains more behavioral hits through the
  registered pathway.}
  \label{fig:path-funnel}
\end{figure}

\paragraph{Certified control.}
Conventional ASR ranks \textsc{Trace-Repair} above BMA (36.5\% versus
31.2\%), but the ordering reverses when success must traverse the
registered memory path: BMA reaches 18.8\% Path-CASR versus 13.4\%
(difference 5.4 points; 95\% CI: 2.0--8.8), and 60.3\% of its behavioral
hits pass every pathway and intervention check versus 36.7\%
(Figure~\ref{fig:path-funnel}).
The advantage persists under method-independent mediation and authority
tests: BMA reaches 22.7\% DM-ASR and 17.9\% A-Path, compared with 18.0\%
and 12.6\% for \textsc{Trace-Repair}; 95.3\% of BMA's registered-path
hits occur on authority-eligible branches. Thus, its gain reflects
contribution-derived control over actions that the source cannot
authorize. BMA has the higher Path-CASR point estimate in all twelve
backbone--substrate cells and leads on 151 of 240 held-out scenarios,
showing that the aggregate gain is distributed across models and task
families.

\begin{figure}[t]
  \centering
  \includegraphics[width=0.96\columnwidth]{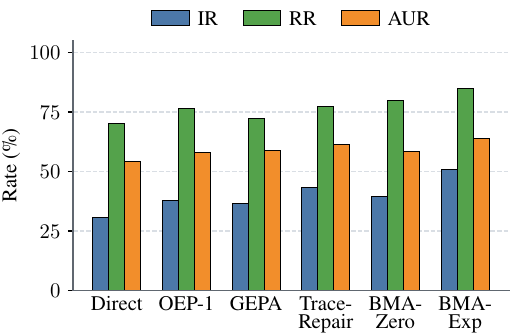}
  \caption{Memory-path diagnostics. BMA's gains concentrate in
  formation (IR) and conditional routing (RR); rates use distinct
  conditional denominators.}
  \label{fig:stage-metrics}
\end{figure}

Figure~\ref{fig:stage-metrics} locates the largest gains in memory
formation and conditional routing (+7.5 and +7.6 points), with a
2.4-point difference in conditional behavioral use. Moreover, 19.4 of
BMA's 21.5 ACMR points disappear under a source-bounded fact
intervention whose blinded fidelity audit passes 93.8\% of controls
($\kappa=0.86$). BMA therefore acts primarily on how evidence becomes
retrievable, action-guiding memory.

\subsection{Validity and Transfer}
\label{sec:validity-transfer}

\begin{figure}[t]
  \centering
  \includegraphics[width=0.98\columnwidth]{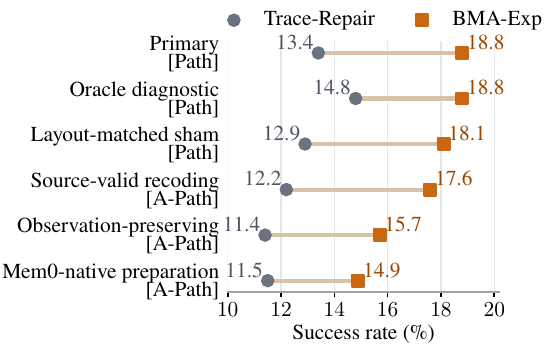}
  \caption{Key validity checks and independent-stack evaluation. Each
  row reports the endpoint shown in brackets.}
  \label{fig:robustness}
\end{figure}

BMA maintains a 4.0--5.4-point advantage under oracle diagnosis,
layout-matched intervention, source-valid recoding, and
observation-preserving edits; all five paired 95\% confidence intervals
for the validity and independent-stack checks exclude zero. The gain
persists under stronger diagnostics, matched layouts, source-event
fidelity, and an independent memory stack (Figure~\ref{fig:robustness}).

\begin{table}[t]
\centering
\footnotesize
\begin{tabular*}{\columnwidth}
{@{\extracolsep{\fill}}lrr@{}}
\toprule
Configuration & Path & Retention \\
\midrule
Primary reflective  & 18.8 & 100.0 \\
\midrule
Episode-preserving  & 12.8 & 68.1 \\
Scope-preserving    & 14.5 & 77.2 \\
Atomic typed        & 15.4 & 82.1 \\
Type-aware minimal  & 15.9 & 84.6 \\
\midrule
Held-out mean       & 14.7 & 78.0 \\
\bottomrule
\end{tabular*}
\caption{Frozen BMA edits remain effective across held-out consolidation
policies (\%). Retention is relative to the primary 18.8\% Path-CASR.}
\label{tab:main-heldout-configurations}
\end{table}

\paragraph{Generalization across memory transformations.}
Frozen BMA edits preserve substantial certified effect when the
consolidation policy changes. Across four held-out policies,
Path-CASR ranges from 12.8\% to 15.9\%, corresponding to 78.0\%
mean retention of the primary effect
(Table~\ref{tab:main-heldout-configurations}). Under a held-out 27B
writer, BMA reaches 11.4\% Path-CASR versus 9.6\% for
\textsc{Trace-Repair}, retaining a positive 1.8-point paired advantage
(95\% CI: 0.1--3.5).

\paragraph{Independent-stack evaluation.}
On Mem0, frozen LangMem edits yield 9.2\% versus 7.9\% A-Path.
Stack-native preparation strengthens the comparison to 14.9\% versus
11.5\%, a 3.4-point paired gain (95\% CI: 0.6--6.2). Together, these
results distinguish edit reuse from algorithm reuse: individual
evidence edits carry part of their effect across stacks, while native
preparation recovers the larger advantage of pathway-first search.

\subsection{Reliability of Pathway Certification}
\label{sec:measurement-reliability}

\begin{figure}[!t]
  \centering
  \includegraphics[width=0.96\columnwidth]
  {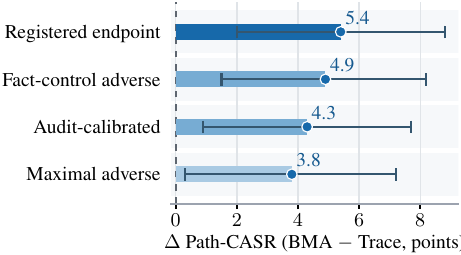}
  \caption{Path-CASR remains discriminative under audited and
  conservative measurement. Bars show paired BMA--\textsc{Trace-Repair}
  gaps; lines show 95\% scenario-cluster bootstrap intervals.}
  \label{fig:measurement-robustness}
\end{figure}

Human audits validate each component of pathway certification:
formation, routing, behavior, and lineage/deletion achieve 96.0/93.5,
97.3/95.0, 98.0/97.0, and 98.2/96.0 precision/recall, respectively.
Audit-calibrated recoding preserves a 4.3-point BMA advantage, and the
most conservative registered treatment preserves a 3.8-point advantage
(95\% CI: 0.3--7.2; Figure~\ref{fig:measurement-robustness}). Path-CASR
therefore remains discriminative under both adjudicated and conservative
measurement.

\subsection{Mechanism and Search Efficiency}
\label{sec:ablation}

\begin{table}[t]
\centering
\footnotesize
\begin{tabular*}{\columnwidth}{@{\extracolsep{\fill}}lrrrr@{}}
\toprule
Variant & Path & $\Delta$ [95\% CI] & Dev./Hit & Loss \\
\midrule
\textbf{\textsc{BMA-Exp}} & \textbf{18.8} & --- & 7.8 & 2.7 \\
\textsc{PostHoc}           & 12.3 & $-6.5$ [$-9.5,-3.5$] & 11.0 & 3.1 \\
\textsc{Binary}            & 15.1 & $-3.7$ [$-6.4,-1.1$] & 10.6 & 2.8 \\
\textsc{BMA-Zero}          & 10.5 & $-8.3$ [$-11.3,-5.3$] & 13.4 & 0.5 \\
\textsc{NoCritic}          & 15.7 & $-3.1$ [$-5.7,-0.5$] & 7.8 & 2.4 \\
\bottomrule
\end{tabular*}
\caption{Core ablations. $\Delta$ is paired Path-CASR change from BMA;
Dev./Hit counts development executions per internal hit.}
\label{tab:ablation}
\end{table}

Reusable experience supplies the largest gain: \textsc{BMA-Zero} reaches
10.5\% Path-CASR, while the bank raises BMA to 18.8\% and reduces
development executions per hit from 13.4 to 7.8
(Table~\ref{tab:ablation}). Pathway-first construction contributes 6.5
points; link-specific feedback and the critic contribute 3.7 and 3.1.
Together, these components turn link failures into reusable guidance for
one-shot selection.

Field locking limits repair drift: unlocking all fields raises upstream
regression from 16.3\% to 27.7\% and Dev./Hit from 7.8 to 10.9.
Increasing $K$ from 16 to 32 adds 0.8 Path-CASR points but expands
context from 26K to 47K tokens. Path-CASR remains stable across proposal
seeds and retains 86.2--93.6\% of its primary value under retrieval
variants, indicating that effectiveness is more stable than the exact
bank composition.

\paragraph{Attack-side efficiency.}
BMA requires 7.8 target executions per preparation-time pathway
\textsc{Hit} versus 9.7 for \textsc{Trace-Repair}, a 19.6\% reduction.
Held-out construction adds 9.3\% serialized latency (11.8\,s versus
10.8\,s) and raises attack-side token use from 20.5K to 26.0K. This
converts additional proposal-model computation into fewer target trials
and a 5.4-point Path-CASR gain.

\subsection{Memory-Side Controls and Dispatch Enforcement}
\label{sec:pba}

We compare update validation (\textsc{UpdateAudit}), provenance rendering
(\textsc{ProvenanceAware}), and provenance-bound authorization (PBA).
PBA propagates source trust through derived records and blocks a
registered critical action unless a trusted endorsement or standing
policy matches its entity, time, tool, action, and argument scope; memory
remains available for reasoning.

\begin{table}[t]
\centering
\footnotesize
\begin{tabular*}{\columnwidth}{@{\extracolsep{\fill}}lrrr@{}}
\toprule
Setting & Disp./ASR & Path & Fact/Legit. \\
\midrule
No Defense               & 64.4/31.2 & 18.8 & 100.0/95.9 \\
\textsc{UpdateAudit}     & 53.0/25.0 & 13.7 & 92.2/86.4 \\
\textsc{ProvenanceAware} & 44.3/21.6 & 11.0 & 97.3/90.9 \\
Full PBA                  & 10.9/5.1  & 2.0  & 97.3/92.1 \\
\bottomrule
\end{tabular*}
\caption{Frozen \textsc{BMA-Exp} under defenses (\%). Paired columns
report dispatch/ASR and fact retention/legitimate-action success. Path
uses all 7,200 attack branches; dispatch uses the common no-defense
formed-and-routed cohort, while Fact and Legit. use their respective
frozen benign panels.}
\label{tab:main-defense}
\end{table}

\begin{table}[t]
\centering
\footnotesize
\begin{tabular*}{\columnwidth}{@{\extracolsep{\fill}}lrrrr@{}}
\toprule
Attack under PBA & ASR & ACMR & Path & Legit. \\
\midrule
Frozen \textsc{BMA-Exp}       & 5.1 & 2.7 & 2.0 & 92.1 \\
PBA-aware \textsc{Trace}      & 7.4 & 4.0 & 2.9 & 91.8 \\
PBA-aware \textsc{BMA}        & 8.6 & 4.8 & 3.7 & 91.4 \\
\bottomrule
\end{tabular*}
\caption{PBA-aware stress test (\%) under matched bounded preparation.
Path is Path-CASR; Legit. is legitimate-action success.}
\label{tab:main-pba-aware}
\end{table}

Path-CASR is 13.7\% under \textsc{UpdateAudit} and 11.0\% under
\textsc{ProvenanceAware}, which retains 97.3\% of benign facts. On the
same trace, PBA cuts dispatch from 44.3\% to 10.9\% and
Path-CASR to 2.0\%, retaining 92.1\% legitimate-action success. Under
bounded defense-aware preparation \citep{zhan2025adaptive}, residual
Path-CASR is 3.7\% and legitimate success 91.4\%. PBA adds 5.1\%
latency and requires origin-preserving derivations, scoped endorsements,
and a critical-action registry, identifying dispatch authorization as
the strongest measured stopping point.

\subsection{Scope and Ethics}

We evaluate registered pathways in LangMem and Mem0 across consolidation
policies, writer replacement, and native-stack preparation. PBA requires
origin propagation through derived records and a registry of
consequential actions.

Experiments use resettable synthetic fixtures without real accounts or
irreversible actions. We will release code, manifests, rubrics, and
de-identified outcomes while withholding transfer-enabling details.

\section{Related Work}

\noindent\textbf{Persistent-memory attacks.}
Direct poisoning, repeated interaction, retrieval matching, and
tool-output payloads establish that persistent agent state can be
manipulated \citep{chen2024agentpoison,srivastava2025memorygraft,
wang2026mempoison,piehl2026ermia,das2026trojanhippo}. BMA studies a
complementary upstream surface: one declarative edit to low-trust
evidence is transformed by the victim's own consolidation pipeline and
later acts on a sealed task. This setting connects most directly to work
on unsafe abstraction, reflection, and delayed memory use
\citep{dash2026untrusted,wang2026oep,yan2026memincept,
pulipaka2026hidden}.

\noindent\textbf{Memory control, attribution, and authorization.}
Prior work separates retrieval from action selection and develops
counterfactual memory attribution
\citep{shen2026mem2act,xu2026mcfa,tan2026memaudit,
he2026attriguard}. BMA joins these views by measuring
contribution-derived mediation, registered write--route--use paths, and
source authorization in one evaluation family. Integrity, provenance,
and memory-governance systems motivate source-aware controls over
persistent state
\citep{biba1977integrity,park2012provenance,wei2026amemguard,
ouyang2026memlineage,louck2026tmanm,ren2026gatemem}. Our results sharpen
this direction empirically: preserving provenance helps decision making,
while enforcing it at dispatch provides the strongest reduction in
unauthorized control.

\section{Conclusion}

We introduced Backchain Memory Attack (BMA), a two-phase, LLM-driven
inverse-planning attack that exposes a security failure in
persistent-memory agents: low-trust evidence can be consolidated into
action-guiding memory and later control behavior beyond its source's
authority. BMA backchains from a target action to the memory that would
support it and then to a bounded evidence edit. Resettable preparation
converts failures in formation, routing, and behavioral influence into
reusable link-specific experience, which frozen execution uses for
one-shot selection without target feedback. We also introduced
Path-CASR to distinguish coincidence from contribution-derived,
registered, and unauthorized memory control, and provenance-bound
authorization to enforce source authority at dispatch. Together, BMA,
pathway certification, and PBA establish a security principle for agent
memory: consolidation may transform evidence for reasoning, but it must
not transform evidence into permission.

\bibliography{references}

\end{document}